\documentclass[twocolumn,aps,superscriptaddress]{revtex4}
\usepackage{amsmath}
\usepackage{graphicx}
\begin{document}
\title{On the electrostatic potential profile in biased molecular wires}
\author{Abraham Nitzan}
\author{Michael Galperin}
\affiliation{School of Chemistry, The Sackler Faculty of Science, Tel Aviv
University, Tel Aviv, 69978, Israel}
\author{Gert-Ludwig Ingold}
\affiliation{Institut f\"ur Physik, Universit\"at Augsburg,
Universit\"atsstra{\ss}e 1, D-86135 Augsburg, Germany}
\author{Hermann Grabert}
\affiliation{Fakult\"at f\"ur Physik, Albert-Ludwigs-Universit\"at,
Hermann-Herder-Stra{\ss}e 3, D-79104 Freiburg, Germany}
\begin{abstract}
The potential profile across a biased molecular junction is calculated within 
the framework of a simple Thomas--Fermi type screening model. In particular, 
the relationship between this profile and the lateral molecular cross section 
is examined. We find that a transition from a linear potential profile to a 
potential that drops mainly near the molecule-metal contacts occurs with 
increasing cross section width, in agreement with numerical quantum 
calculations.
\end{abstract}
\maketitle

\section{Introduction}
Molecular wires are molecules that bridge between metallic leads to form a 
nano-conductor whose current-voltage characteristic reflects the electronic 
structure of the molecule. The conductance may be controlled via its 
dependence on molecular properties. Equally important is the use of such 
molecular junctions as capacitive elements in nano-electronics. 

Understanding the behavior of such junctions under potential bias is a 
prerequisite for elucidating their transport properties. The importance of this 
issue for the conductance of molecular junctions was recently emphasized by 
Datta and coworkers \cite{datta97,tian98}, who have shown, within a simple 
Extended-H{\"u}ckel (EH) model for $\alpha$,$\alpha'$-xylyl dithiol bridging 
between two gold leads, that the potential profile (imposed on the molecule as 
input to the EH calculation) had a profound effect on quantitative as well as 
qualitative aspects of the calculated current-voltage characteristic. The best 
fit to experimental results was obtained from a model that assumed (a) a flat 
potential profile in the interior of the molecular bridge, i.e. the potential 
drop occurs only at the molecule-lead contacts and (b) a symmetric distribution 
of the potential drop at the two contacts, i.e. for a total voltage $\Phi$ the 
potential drops at each molecular edge by $\Phi/2$. 

This picture is supported by a recent model calculation by Mujica 
\textit{et al.}\ \cite{mujic00}, where the Schr{\"o}dinger equation (on the 
Hartree level) was solved in conjunction with the Poisson equation to yield 
both the electronic structure and the spatial distribution of the electrostatic 
potential \cite{hiros95}. It was found that beyond a screening distance of the 
order of 1--3 atomic lengths the potential is flat along the model molecular 
chain. 

Ab initio calculations with open system boundary conditions reveal a different 
picture: Lang and Avouris \cite{lang00} have found for a chain of seven carbon 
atoms connecting between jellium leads that a substantial part of the voltage 
drop occurs along the carbon chain itself. Damle 
\textit{et al.}\ \cite{damle01} have obtained similar results for a chain of 
gold atoms as well as for a molecular wire --- phenyl-dithiol bridging between 
two gold electrodes \cite{footn}. In an earlier work, Pernas 
\textit{et al.}\ \cite{perna90} have determined that the potential along a 
model molecular wire is flat in the absence of scattering centers, however 
these authors have derived the local potential from a questionable local charge 
neutrality condition. 

Recently, Weber \textit{et al.}\ \cite{weber02} have considered the voltage
profile across 9,10-Bis((2'-para-mercaptophenyl)-ethinyl)-anthracene coupled to 
two Au$_{29}$ clusters. Their density functional theory calculations thus go 
beyond the assumption of a structureless metallic electrode and take into 
account the specific properties of the bond between the molecule and the gold 
atom in its local environment. 

On the experimental side, Bachtold \textit{et al.}\ \cite{bacht00} have used 
scanned probe microscopy of both single-walled and multi-walled carbon 
nanotubes (SWNT and MWNT, respectively) to measure the potential drop along 
such nanotubes connecting between two gold electrodes. They find an 
approximately linear drop of the potential in a MWNT of diameter 9nm while for 
a metallic SWNT bundle of diameter 2.5nm the potential is flat beyond the 
screening regions at the tube edges. It should be emphasized that these 
experiments cannot be related directly to the calculations discussed above. The 
nanotube length is a few microns and impurity and defect scattering may be 
effective as is most certainly the case in the MWNT measurement. The flat 
potential seen in the metallic SWNT measurement is in fact a remarkable 
observation implying a very long mean free path ($>1\mu$m) for electrons in 
these room temperature structures.

It is clear from the studies described above that while the computational 
methodology for evaluating the potential distribution on a biased molecular 
wire is available, a complete understanding of the way this distribution 
behaves in different wires is lacking. In this respect simple models that focus 
on generic properties of conducting constrictions including molecular wires are 
useful. The calculations of Pernas \textit{et al.} \cite{perna90} provide such 
a model that is however hampered, as already stated by the restriction of local 
charge neutrality. The calculation of Mujica \textit{et al.} \cite{mujic00} is 
also based on a generic molecular model, however, by using a 1-dimensional 
Poisson equation for the electrostatic potential these authors tacitly assume 
a molecular bridge whose lateral dimension is much larger than the screening 
length. In view of the fact that the width of molecular wires is often just a 
few angstroms, such an assumption is overly restrictive. Clearly, the 
magnitudes of the lateral width of the wire and the screening length should be 
important generic quantities for this issue. In this paper we present a simple 
model calculation that takes the relative magnitudes of these variables 
explicitly into account. We describe the model in Section \ref{sec:model}, 
present the calculation in Section \ref{sec:potential} and some results and 
discussion in Section \ref{sec:results}.

\section{The Model}
\label{sec:model}

\begin{figure}
\includegraphics[width=0.9\linewidth]{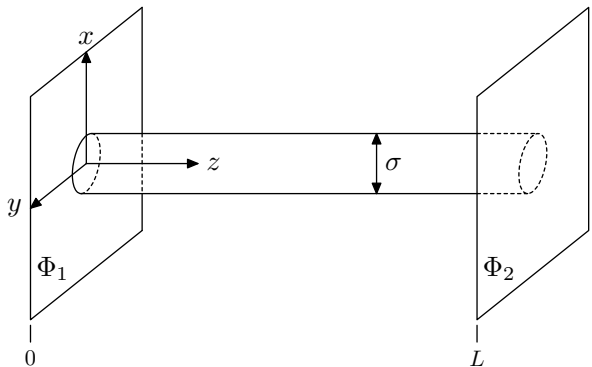}
\caption{The general setup contains a molecular wire modeled as a cylinder
of length $L$ and diameter $\sigma$ between two perfect conductors with
potentials $\Phi_1$ and $\Phi_2$.}
\label{fig:geometry}
\end{figure}

The molecular wire is modeled as a cylinder of length $L$ and diameter of order 
$\sigma$ (the exact way in which $\sigma$ enters into the model calculation is 
explained below), perpendicular to and connecting between two planar metal 
electrode surfaces. As depicted in Fig.~\ref{fig:geometry}, the cylinder is 
oriented parallel to the $z$ axis, with its axis going through the origin in 
the $xy$ plane. The two electrodes are assumed to be ideal conductors
implying a constant potential on the entire surface of each electrode.
We set the potentials at the left and right wire-electrode 
interface to be $\Phi_1=\Delta/2$ and $\Phi_2=-\Delta/2$, respectively.
In view of (\ref{eq:screening}) below, this guarantees a vanishing mean
potential in $z$-direction and thus a neutral molecule. Finally, we
restrict the discussion of the potential profile to blocking junctions
between electrodes and molecule so that no current is flowing.

As in Ref.~\cite{mujic00} we assume that the wire material does have screening 
capacity, and is characterized by a screening length $\lambda$. 
It should be noted that the existence, nature and size of such screening length 
in molecular systems is an open question with probably non-unique answer. 
Molecules with large gaps between their highest occupied and lowest unoccupied 
molecular orbitals (HOMO and LUMO) will obviously screen poorly, while highly 
conjugated molecules with low HOMO-LUMO gap will screen relatively well. 

In the present discussion we assume that over the relevant length scales (of 
order $L$) screening is described by a Poisson equation
\begin{equation}
\nabla^2\Phi = -4\pi\rho\;.
\label{eq:poisson}
\end{equation}
According to the cylinder symmetry of the molecule, the charge density
$\rho(r_{\parallel},z)$ depends on the radial distance $r_{\parallel}$ from
the wire axis and the position $z$ along the wire. In transversal direction, 
the charge density is assumed to be determined by a given molecular
electron distribution represented by a factor $F(r_{\parallel})$. The
longitudinal part depends on the potential along the molecular axis. The
screening is then described by
\begin{equation}
4\pi\rho(r_{\parallel},z) = -\frac{1}{\lambda^2}F(r_{\parallel})\Phi(0,z)
\label{eq:screening}
\end{equation}
which together with (\ref{eq:poisson}) will allow us to determine the potential
profile.

Any assumption about the functional form of $F(r_{\parallel})$ is in fact an 
assumption about the confinement of the molecular charge distribution in the 
molecular cylinder and in our generic model it is sufficient to take a form 
that reflects the molecular thickness $\sigma$. Other details of
$F(r_{\parallel})$ are expected to be of secondary importance. 

In the three-dimensional Thomas Fermi model for screening in a gas of
electrons with charge $e$ and mass $m_{\rm e}$, the screening length $\lambda$ 
of (\ref{eq:screening}) is related to the electron density $n$ by
\begin{equation}
\lambda = \left(\frac{E_{\rm F}}{6\pi ne^2}\right)^{1/2}
\end{equation}
with the Fermi energy
\begin{equation}
E_{\rm F} = \frac{(3\pi^2 n)^{2/3}\hbar^2}{2m_{\rm e}}\,.
\end{equation}
At metallic electron densities $\lambda$ is typically of the order of
1\,\AA. To have efficient 
screening in a molecular system electrons (or holes) must occupy molecular 
states that are effectively delocalized over the length of the bridge. Charge 
doping by transfer from the metal electrode to the molecular bridge may be one 
source of such electrons. Their density is expected to be considerably lower 
than metallic, implying a larger characteristic screening length. We expect 
that a calculation based on (\ref{eq:poisson}) and (\ref{eq:screening}) 
that uses metallic electron density to estimate $\lambda$ will provide an upper 
bound on the effective screening in a molecular wire.
	
\section{The potential distribution}
\label{sec:potential}

Using the model described in the previous section, our problem is to solve the 
equation
\begin{equation}
\nabla^2\Phi(r_{\parallel},z) = \frac{1}{\lambda^2} F(r_{\parallel})\Phi(0,z)
\end{equation}
in the range $0\le z\le L$ subject to the boundary conditions
\begin{equation}
\Phi(r_{\parallel},0) = \Delta/2\quad\mbox{and}\quad
\Phi(r_{\parallel},L) = -\Delta/2\;.
\end{equation}
It is convenient to decompose the full potential 
\begin{equation}
\Phi(r_{\parallel},z)=\Phi_0(z) + \phi(r_{\parallel},z)
\end{equation}
into a first term describing the bare potential
\begin{equation}
\Phi_0(z)=\Delta\left(\frac{1}{2}-\frac{z}{L}\right)
\label{eq:phi0}
\end{equation}
in the absence of a molecule and a second term which reflects the additional 
potential $\phi(r_{\parallel},z)$ satisfying the boundary conditions 
$\phi(r_{\parallel},0)=\phi(r_{\parallel},L)=0$.

The resulting differential equation 
\begin{equation}
\nabla^2\phi(r_{\parallel},z) =\frac{1}{\lambda^2}F(r_{\parallel})
\left[\Phi_0(z)+\phi(0,z)\right]
\end{equation}
may be solved by the Fourier ansatz
\begin{equation}
\phi(r_{\parallel},z) = \int\frac{{\rm d}^2k_{\parallel}}{(2\pi)^2}
{\rm e}^{{\rm i}{\bf k}_{\parallel}\cdot{\bf r}_{\parallel}}\sum_{n=1}^{\infty}
\hat\phi_n(k_{\parallel})\sin\!\left(\frac{\pi n}{L}z\right)\;.
\end{equation}
After expressing the bare potential profile in terms of a Fourier series
one arrives at
\begin{equation}
\begin{aligned}
\hat\phi_n(k_{\parallel}) &= -\frac{1}{\lambda^2[k_{\parallel}^2+\displaystyle
(\pi n/L)^2]}\hat F(k_{\parallel})\\
&\qquad\times\left[\Delta\frac{1+(-1)^n}{\pi n} + \int\frac{d^2k_{\parallel}'}
{(2\pi)^2}\hat\phi_n(k_{\parallel}')\right]
\end{aligned}
\label{eq:phift}
\end{equation}
where
\begin{equation}
\hat F(k_{\parallel}) = \int{\rm d}^2r_{\parallel}{\rm e}^{-{\rm i}
{\bf k}_{\parallel}\cdot{\bf r}_{\parallel}}F(r_{\parallel})\;.
\end{equation}
For the potential profile along the molecular axis, only the transversal 
integral over the Fourier coefficients $\hat\phi_n(k_{\parallel})$ is
needed which may easily be obtained from (\ref{eq:phift}). Due to the
symmetry of the bare potential only even Fourier coefficients are found to 
contribute. We thus arrive at our main result describing the potential profile 
along the molecule 
\begin{equation}
\Phi(0,z) = \Phi_0(z)-\frac{\Delta}{\pi}\sum_{n=1}^{\infty}
\frac{F_n}{n(1+F_n)}\sin\!\left(\frac{2\pi n}{L}z\right)\;.
\label{eq:phi00zf}
\end{equation}
The coefficients $F_n$ accounting for the influence of screening are given
by
\begin{equation}
\begin{aligned}
F_n&=\frac{1}{\lambda^2}\int\frac{{\rm d}^2k_{\parallel}}{(2\pi)^2}
\frac{\hat F(k_{\parallel})}{k_{\parallel}^2+(2\pi n/L)^2}\\
&=\frac{1}{\lambda^2}\int_0^{\infty}{\rm d}r_{\parallel} r_{\parallel}
F(r_{\parallel})K_0\!\left(\frac{2\pi n}{L}r_{\parallel}\right)
\label{eq:fn}
\end{aligned}
\end{equation}
where $K_0$ denotes a modified Bessel function. In the limit of very small
screening length, $\lambda\to0$, it is possible to show by evaluating the sum 
in (\ref{eq:phi00zf}) that the potential along the wire vanishes and the
entire voltage drop occurs at the interface with the electrodes.

For the following discussion, it is convenient to introduce a measure of
the deviation of the voltage profile $\Phi(0,z)$ from the linear behavior
(\ref{eq:phi0}). Since the integral over $\Phi(0,z)-\Phi_0(z)$ vanishes
for a neutral molecule, we use instead
\begin{equation}
\delta = \left[\frac{12}{\Delta^2L}\int_0^L{\rm d}z\big(\Phi(0,z)-
\Phi_0(z)\big)^2\right]^{1/2}\;.
\label{eq:deltadef}
\end{equation}
This quantity is normalized such that it equals 1 if the voltage drop
occurs entirely at the ends of the molecule while it vanishes for a linear 
potential profile. Employing (\ref{eq:phi00zf}), one may express $\delta$ in 
terms of the coefficients defined by (\ref{eq:fn}) as
\begin{equation}
\delta = \frac{6^{1/2}}{\pi}\left[\sum_{n=1}^{\infty}
\frac{F_n^2}{n^2(1+F_n)^2}\right]^{1/2}\;.
\label{eq:delta}
\end{equation}

\section{Results and discussion}
\label{sec:results}

\begin{figure}
\includegraphics[width=0.9\linewidth]{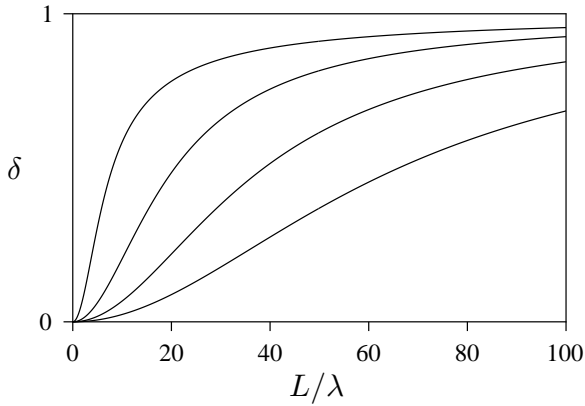}
\caption{The deviation $\delta$ (cf.\ (\protect\ref{eq:deltadef})) of the 
voltage profile from the linear behavior is shown as a function of the ratio of 
wire length $L$ and screening length $\lambda$. The four curves correspond to 
$\sigma\to\infty$ (cf.\ (\protect\ref{eq:deltainf})), $\sigma/L = 0.05, 0.02,$ 
and $0.01$ from the upper to the lower curve.}
\label{fig:delta}
\end{figure}

We now address the dependence of the potential profile on the width of the 
molecular wire and start with the limiting case of an infinitely thick
molecule or, equivalently, a large number of molecules in parallel present
between the two electrodes. Then, $F(r_{\parallel})=1$ and one finds 
from (\ref{eq:fn}) 
\begin{equation}
F_n=\left(\frac{L}{2\pi\lambda n}\right)^2\;. 
\label{eq:fninf}
\end{equation}
Using
\begin{equation}
\sum_{n=1}^{\infty}\frac{\sin(nx)}{n(n^2+a^2)} =
\frac{\pi}{2a^2}\frac{\sinh\big(a(x-\pi)\big)}{\sinh(a\pi)}
-\frac{x-\pi}{2a^2}\;,
\end{equation}
(\ref{eq:phi00zf}) yields for the potential profile
\begin{equation}
\Phi(0,z) = \frac{\Delta}{2}\frac{\sinh\!\left(\displaystyle
\frac{L-2z}{2\lambda}\right)}{\sinh(L/2\lambda)}\;.
\end{equation}
The deviation from the linear voltage drop can be quantified by inserting
(\ref{eq:fninf}) into (\ref{eq:delta}). Evaluating the sum, one finds
\begin{equation}
\delta = \left[1+24\frac{\lambda^2}{L^2}-9\frac{\lambda}{L}\coth\left(
\frac{L}{2\lambda}\right)-\frac{3}{2}\frac{1}{\sinh(L/2\lambda)}\right]^{1/2}\;.
\label{eq:deltainf}
\end{equation}
This result is shown in Fig.~\ref{fig:delta} as uppermost curve.
In the limit of very large screening length, $\lambda\to\infty$, $\delta$
vanishes, thereby indicating the expected linear voltage drop. On the 
other hand, for very short screening length, $\lambda\to0$, $\delta$ approaches
one and the entire voltage drops at the interfaces between wire and electrode.

For the case of finite width, we employ the Gaussian charge distribution
\begin{equation}
F(r_{\parallel}) = \exp(-r_{\parallel}^2/2\sigma^2)\;.
\label{eq:gaussian}
\end{equation}
This function is not normalized and therefore describes a charge distribution
with a density in the center independent of the width $\sigma$. Such a 
situation arises when the diameter of a molecular layer can be controlled. 
Then, the charge density in the center appears in the screening length 
$\lambda$. In contrast, in the somewhat unrealistic case where the charge 
density on the wire is changed, the function $F(r_{\parallel})$ would have to 
be normalized.

One advantage of the Gaussian distribution (\ref{eq:gaussian}) is the fact that 
the coefficients (\ref{eq:fn}) may still be expressed analytically in terms of 
an exponential integral
\begin{equation}
F_n=\frac{1}{2}\left(\frac{\sigma}{\lambda}\right)^2 
{\rm e}^{\chi}\int_{\chi}^{\infty}{\rm d}u
\frac{{\rm e}^{-u}}{u}
\label{eq:fnf}
\end{equation}
with
\begin{equation}
\chi=\frac{1}{2}\left(2\pi n\frac{\sigma}{L}\right)^2\;.
\end{equation}
With this result the potential profile can be evaluated numerically according
to (\ref{eq:phi00zf}) while the deviation from the linear voltage drop is
obtained from (\ref{eq:delta}).

In Fig.~\ref{fig:delta}, the deviation $\delta$ of the voltage profile from
the linear behavior (\ref{eq:phi0}) is shown for different values of the
wire thickness $\sigma$. The uppermost curve corresponds to the limit of
a thick molecular layer $\sigma\to\infty$ which was discussed above. The three
other curves correspond to $\sigma/L=0.05, 0.02,$ and $0.01$ from top to bottom.
As these results demonstrate, a reduction of $\sigma$ causes a reduction
of $\delta$ indicating that the voltage profile approaches the linear voltage
drop. This behavior can be understood in terms of a reduction of screening due 
to the reduced molecular layer. However, this effect becomes only relevant
for $\sigma\ll L$. As discussed above, the limit $\lambda\to0$ leads to a
constant potential along the molecular wire. Therefore, all curves shown
in Fig.~\ref{fig:delta} tend to $\delta=1$ in this limit even though this
is not shown in the figure.

\begin{figure}
\includegraphics[width=\linewidth]{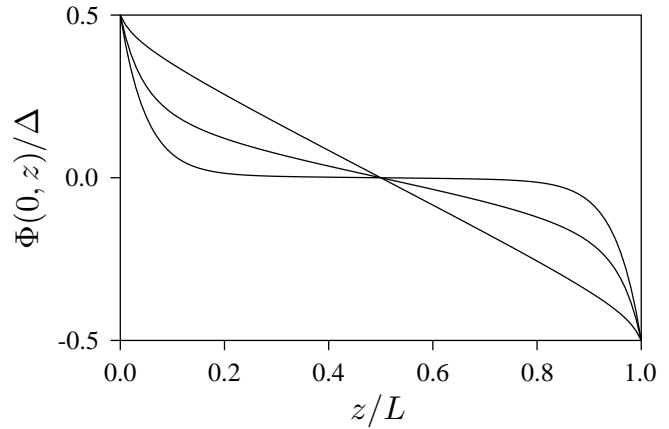}
\caption{The potential profile along a molecular wire computed from 
(\ref{eq:phi00zf}) and (\ref{eq:fnf}) is shown for a screening length
$\lambda/L=0.05$. The thickness parameter $\sigma/L$ takes the values
$0.0125, 0.05,$ and $0.5$ from the almost linear behavior to the voltage
profile containing almost a plateau.}
\label{fig1}
\end{figure}

\begin{figure}
\includegraphics[width=\linewidth]{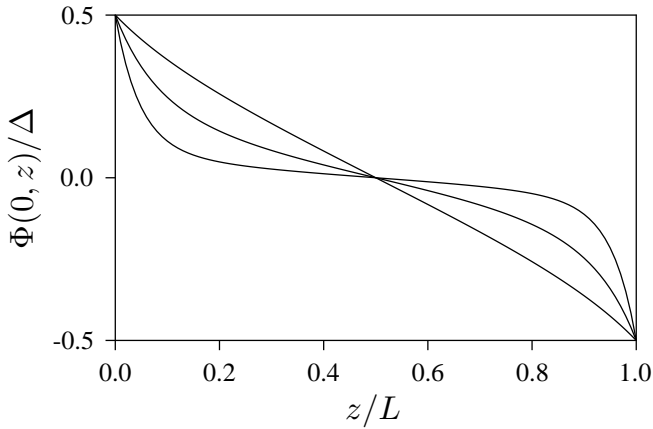}
\caption{Same as Fig.~\protect\ref{fig1} but for fixed thickness parameter
$\sigma/L=0.125$ and varying screening length $\lambda/L=0.25, 0.1,$ and
$0.05$ from the almost linear behavior to the voltage profile containing
almost a plateau.}
\label{fig2}
\end{figure}

We now turn to a discussion of the voltage profiles themselves.
Figure~\ref{fig1} depicts results obtained from (\ref{eq:phi00zf}) using 
$F_n$ from (\ref{eq:fnf}) and $\Phi_0(z)$ as defined in (\ref{eq:phi0}).
The dimensionless screening length $\lambda/L=0.05$ implies for a typical
metallic screening length $\lambda=2\,\mbox{a.u.}$ a wire length of 
$L=20\,\mbox{a.u.}$. The thickness parameter $\sigma/L$ for the three
different curves are $0.0125, 0.05,$ and $0.5$, where the voltage profile
becomes more and more linear as $\sigma$ decreases. As already mentioned,
this may be understood in terms of the reduced screening. Fig.~\ref{fig2} shows
similar results for a wire with a ratio between the typical diameter and the
wire length of $\sigma/L=0.125$. Here, the dimensionless screening length
takes the decreasing values $\lambda/L=0.25, 0.1,$ and $0.05$ with increasing
deviation from the linear voltage profile.

\begin{figure}
\includegraphics[width=\linewidth]{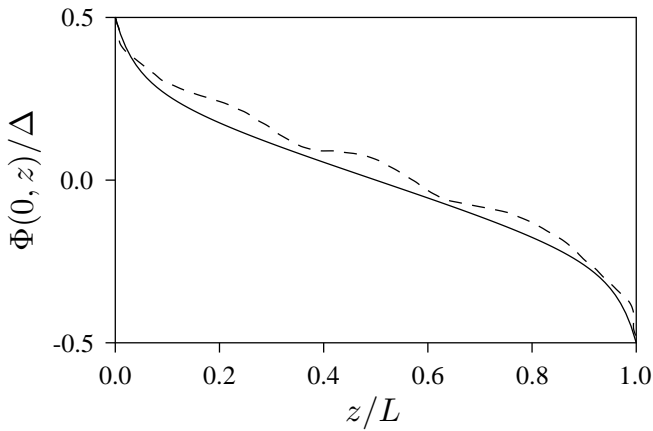}
\caption{The potential drop across a chain of 6 gold atoms placed between two
gold electrodes as obtained from an ab inito calculation in
Ref.~\protect\cite{damle01} (dashed line) is fitted by a voltage profile
(\protect\ref{eq:phi00zf}) depicted as full line. The best fit was obtained
for $\sigma/L=0.032$ and $\lambda/L=0.052$.}
\label{fig3}
\end{figure}

In Fig.~\ref{fig3} we present a fit of our voltage profile to the ab initio 
calculation of Ref.~\cite{damle01}. A least square fit resulted in a
screening length of $\lambda/L=0.052$ and a wire width of $\sigma/L=0.032$.
For a distance $L$ of approximately $34\,\mbox{a.u.}$, this yields the 
reasonable value of $1.1\,\mbox{a.u.}$ for the radial extent of the electron 
density.

The parameters $\sigma/L$ and $\lambda/L$ employed allow for a rather good 
approximation of the results of the ab initio calculation. However, there are 
two significant differences. The Friedel oscillations found in the quantum 
calculation cannot be obtained within our classical approach. In a
tight-binding model description of the molecular wire, Friedel oscillations
naturally arise from a breaking of electron-hole symmetry \cite{pleut02}.
Secondly, the systematic shift between the two voltage profiles in
Fig.~\ref{fig3} indicates that the wire in the ab initio calculation was 
charged while our wire is always assumed to be neutral.

It is obvious from these results that the ratio of the
wire thickness to the screening length does constitute an 
important generic attribute that determines the general behavior of the 
potential bias distribution along a molecular wire. The relatively good fit 
obtained between the model calculations and the ab initio results for a chain 
of gold atoms using reasonable geometric parameters supports this conclusion. 
We note in passing that the flat potential distribution observed \cite{bacht00} 
for metallic single walled carbon nanotubes of thickness $\sim20\,\mbox{a.u.}$ 
is consistent with the results shown in Fig.~\ref{fig1}. One should keep in 
mind however that apart from its intrinsic simplicity, the model used in this 
work suffers from two important shortcomings. Firstly, the use of a simple 
screening property as described by (\ref{eq:poisson}) and 
(\ref{eq:screening}) cannot be justified for all molecules, and certainly not 
for arbitrary distances. Even when such screening applies, the magnitude of the 
screening parameter $\lambda$ is not known and is expected to depend on the 
amount of equilibrium charge transfer between the wire and the metal leads. 
Secondly, a complete calculation of the potential profile along a molecular 
junction should take into account the fact that some of this drop may take 
place on the metal leads near the junction. Such a behavior was found in the 
calculation of Ref.~\cite{lang00}.

\section{Conclusions}

The potential distribution along molecular wires in biased molecular junctions 
is determined in principle by the detailed electronic structure of the wire and 
by the response of this structure to the molecule-lead contacts and to the 
bias. The present study has identified the wire thickness as one of two 
generic attributes that largely determine the way the potential drops along the 
wire. Increasing this parameter leads to a crossover from a three-dimensional
electrostatic problem to an effectively one-dimensional situation. The
accompanying increase in screening causes a transition from a linear potential
profile to a situation where the potential drops mostly at the interfaces
between wire and electrode. The other, less accessible molecular property is 
its ability to screen a local charge. In the present model calculation we have 
used a simple screening length parameter to model this molecular property, but 
further studies are needed for a better characterization of this property. 

\begin{acknowledgments}
GLI is grateful to S.~Yaliraki and J.~Lehmann for stimulating discussions.
Three of us (AN, GLI, and HG) would like to thank the Institute for Theoretical 
Physics at UCSB for hospitality during the workshop on ``Nanoscience'' where 
this work was started. This research was supported in part by the National 
Science Foundation under Grant No.\ PHY99-07949, by the Israel Science
Foundation, and by the Israel Ministry of Science.
\end{acknowledgments}


\begin{thebibliography}{}
\bibitem{footn}
In these ab initio calculations the potential is calculated from the 
combination of the external field and the Hartree potential due to all 
molecular charges, using an extended molecular model that includes two cluster 
segments of the metal leads.
\end{thebibliography}
\end{document}